\documentclass[aps,prb,twocolumn,groupedaddress,amsmath,amssymb,showpacs]{revtex4-1}

\usepackage{graphicx,graphics,color,epsfig}
\usepackage{amsmath}
\usepackage{amssymb}
\usepackage{amsfonts}
\usepackage{amsfonts}
\usepackage{epstopdf}
\usepackage{bm}
\usepackage{times,xspace}
\usepackage{color}
\usepackage[utf8]{inputenc}
\usepackage[normalem]{ulem}

\def\beq{\begin{eqnarray}}\def\eeq{\end{eqnarray}}
\def\be{\begin{equation}}\def\ee{\end{equation}}
\def\g{\gamma}
\def\r{\rho}
\def\s{\sigma}
\def\m{\mu}

\def\a{\alpha}

\def\b{\beta}
\def\d{\delta}

\def\D{\Delta}

\def\pd{\partial}

\def\ra{\rangle}

\def\w{\omega}

\begin{document}

\title{Assembling topological insulators with lasers}

\author{Sayonee Ray$^{1}$\footnote{sayoneeiisc@gmail.com}, Kallol Sen$^{2}$\footnote{kallol@cts.iisc.ernet.in} and Tanmoy Das$^{1}$\footnote{tnmydas@gmail.com}\\
{$^{1}$Department of Physics, Indian Institute of Science, Bangalore-560012, India.\\
$^{2}$Center for High Energy Physics, Indian Institute of Science, Bangalore-560012, India.
}}

\date{\today}

\begin{abstract}
Despite the realizations of spin-orbit (SO) coupling and synthetic gauge fields in optical lattices, the associated time-reversal symmetry breaking, and 1D nature of the observed SO coupling pose challenges to obtain intrinsic $Z_2$ topological insulator. We propose here a model optical device for engineering intrinsic $Z_2$ topological insulator which can be easily set up with the existing tools. The device is made of a periodic lattice of quantum mechanically connected atomic wires (dubbed SO wires) in which the laser generated SO coupling ($\alpha_{\bf k}$, with ${\bf k}$ being the momentum) is reversed in every alternating wires as $\pm\alpha_{\bf k}$. The associated small Zeeman terms are also automatically reversed in any two adjacent  SO wires, which allow to effectively restore the global time-reversal (TR) symmetry. Therefore, the two SO wires serve as the TR partner to each other which is an important ingredient for $Z_2$ topological insulators according to the Kane-Mele model. These properties ensure a non-trivial $Z_2$ invariant topological insulator phase with protected edge states. We also discuss that a non-local current measurement can be used to detect the chiral edge states. 
\end{abstract}





\maketitle

Optical lattice provides a model `breadboard' to imprint diverse quantum and topological phases of ultracold fermionic and bosonic atoms.\cite{rev1,rev2,rev3,rev4,Gauge2,Gauge_AB} The realization of the synthetic gauge field in optical lattices,\cite{Gauge,Gauge2,Gauge_AB} analogous to magnetic field in solid state systems, is one of the major triumph in this field. This discovery has opened up possibilities for devising new and exotic quantum and topological phases, some of which may have even no analog with the solid state counterparts. Many exotic properties such as geometric Berry phase, Harper-Hofstadter butterfly,\cite{HHB} spin-orbit coupling (SOC),\cite{lin} time-reversal (TR) symmetry breaking Haldane lattice,\cite{jotzu} quantum spin-Hall insulator (QSH) \cite{Goldman,aidelsburger}, non-trivial edge states,\cite{mancini,Stuhl} are subsequently synthesized in these systems.

\begin{figure}[t]
\centering
\rotatebox[origin=c]{0}{\includegraphics[width=0.99\columnwidth]{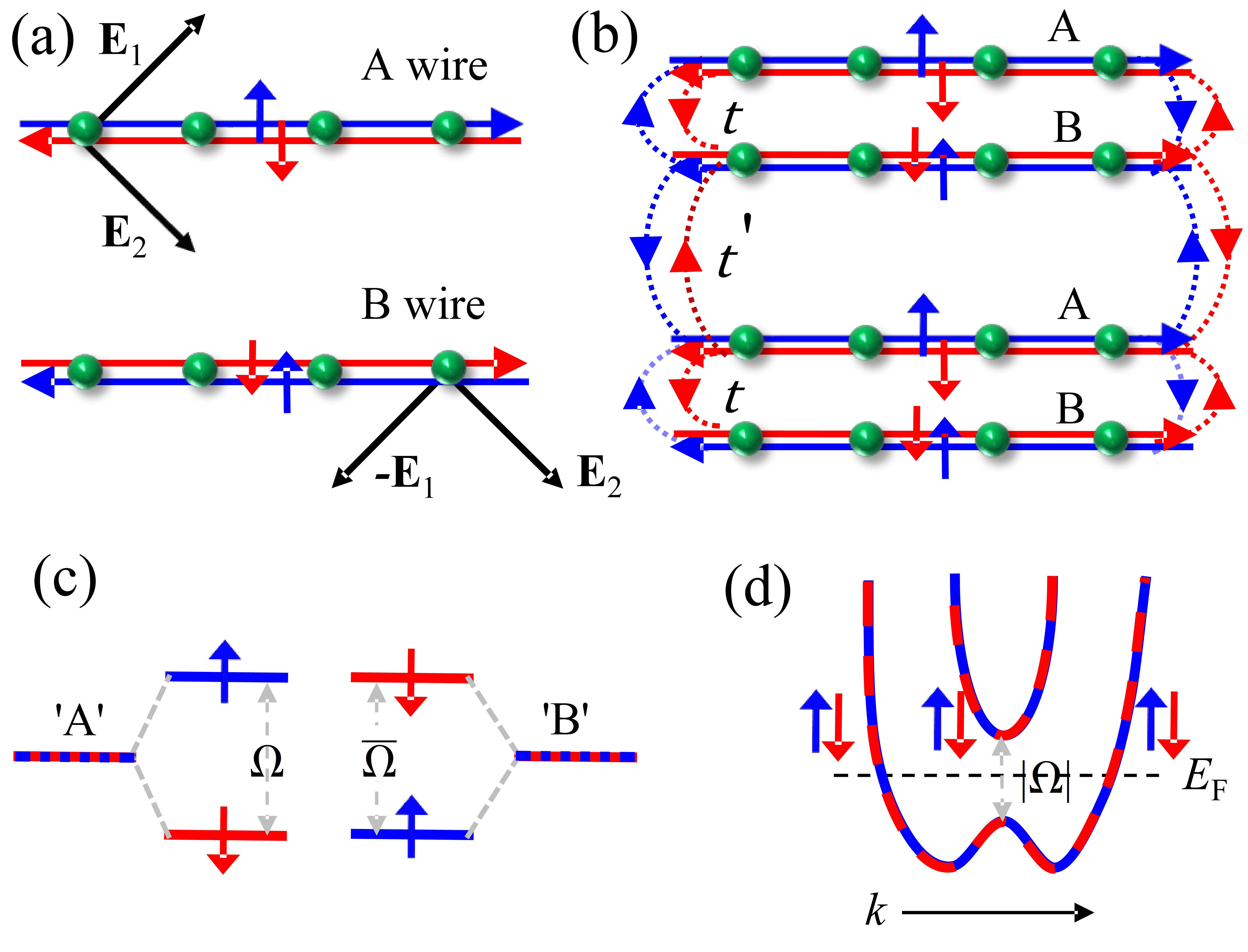}}
\caption{Schematic drawing of the setup. (a) Building block: Top and bottom wires depict the `A' and `B' SO wires of Rb$^{87}$ atoms (periodic along the $x$-axis) with opposite 1D SOCs $\pm\alpha_{\bf k}$. Vertical arrows dictate spins, while horizontal arrows represent their corresponding direction of motion in each wire. (b) Two double SO wires with different nearest-neighbor inter-wire distances aligned along the $y$-axis. Dashed arrows with different colors denote representative inter-wire hoppings for a given spin configuration. $t$ and $t^{\prime}$ are the inter-wire hopping terms. (c) Energy levels for `A' and `B' wires split by opposite Zeeman coupling ($\pm\Omega$). (d) In the corresponding $k$-space, the bands become spin-degenerate with a gap at the $\Gamma$-point without breaking the TR symmetry.
}
\label{fig1}
\end{figure}

Chiral motion of particles, arising from either staggered hopping or SOC, is instrumental to various topological phases of matter\cite{TIreviewCK,TIreviewSCZ,TIreviewTD}. Diverse forms of TR invariant topological and QSH states have been realized in solid state frameworks. Furthermore, a recent realization of SOC with detuned lasers\cite{lin, bliokh} has provided the opportunity that topological phases can also be obtained in optical lattices\cite{aidelsburger}. However a number of shortcomings, intrinsic to the optical lattice frameworks, hinders setting up $Z_2$ topological insulators (TIs) in this framework. For example, due to the inevitable presence of the Zeeman-like term, although often estimated to be small, the TR symmetry becomes inherently broken. Similarly, the SOC in optical devices can be synthesized easily in one-dimensional (1D) atomic chain,\cite{lin} while its generalization to higher dimensions is cumbersome.\cite{SOC2D,SOC2D2} A similar difficulty arises in solids when Rashba- and Dresselhaus-type SOCs possess equal strengths, subsisting only a residual 1D SOC, or in quasi-1D quantum wires. Such 1D SOC prevents electrons to form closed orbit motion in the bulk. Since the localization of counter-propagating `chiral orbits' without breaking the TR symmetry is the key for 2D $Z_2$ TIs, observation of them has proven challenging in optical devices.

We propose an optical device to assemble $Z_2$ TIs in 2D (extension to 3D follows the same principle), as illustrated in Fig.~\ref{fig1}. As two detuned lasers are directed perpendicular to each other, it generates a SOC, with equal Rashba and Dresselhaus strengths, at a 45$^o$ angle from both lasers, say $k_x$ direction\cite{lin}. We take two other lasers of the same configuration, but aligned anti-parallel to the above SO wire in such a way that a SOC commences along the $-k_x$ direction. These two counter-propagating SO wires are referred as `A' and `B' wires as shown in Fig.~\ref{fig1}(a). For engineering advantage, layers of `A' and `B' wires do not need to be on the same plane, and can be assembled on a bi-layer framework.\cite{footnote_setup} If the spin-up atom is right-moving in `A' wire, it becomes left moving in `B' wire and vice versa. Furthermore, as shown in the supplementary material, the Zeeman term for two such wires consequently become completely reversed ($\pm\Omega$), giving opposite spin splittings as shown Fig.~\ref{fig1}(c). Therefore, `A' and `B' wires possess positive and negative band gaps, as in the case of CdTe and HgTe systems, respectively\cite{bhz}. The byproduct of this setup is that as the two wires are brought closer, their combined structure creates an effective spin-degenerate band structure at all $k$-points. Therefore, a band gap can be opened at the TR invariant $k$-points without breaking the TR symmetry as shown in Fig.~\ref{fig1}(d). The effective band gap at the $\Gamma$-point is a combination of the Zeeman term, and the inter-wire (spin conserving) hopping amplitude, $t$. We show below that as the SOC strength exceeds the atom's kinetic energy terms, the valence band in Fig.~\ref{fig1}(d) fails to cross the Fermi level ($E_F$) and an insulating state arises. 

The associated emergence of $Z_2$ topological invariant without any further tuning can be understood from the combinations of band inversion phenomena as proposed in HgTe/CdTe heterostructure,\cite{bhz} and the `TR polarization' as proposed by Kane and Mele.\cite{kane} According to the Kane-Mele formalism,\cite{kane} a $Z_2$ invariant arises if a fermion's wavefunction switches to its TR conjugate odd number of times in traversing half of the Brillouin zone (BZ). This criterion is embedded automatically in our structure since the two SO wires serve as TR partners to each other. As the inter-wire hopping becomes comparable to the intra-wire hopping, the right-moving spin-up atom in `A' wire hops to `B' wire and becomes left-moving, and vice versa. Finally, as the spin up atom hops back to the original `A' wire, it encircles a closed loop, as illustrated in Fig.~\ref{fig1}(b). In this process opposite spin states form counter-helical orbits without breaking the TR symmetry and thus become localized in the bulk. In analogy with the HgTe/CdTe heterostucture,\cite{bhz} the inverted band gap between `A' and `B' wires, Fig.~\ref{fig1}(c), ensures a band inversion at the $\Gamma$-point. We have computed the generalized $Z_2$ invariant for this setup, which holds in both 2D and 3D, and supplemented the results with calculations of edge states. The proposed setup is also applicable in solid state frameworks such as in quasi-1D quantum wires, where SOC and electronic properties are largely tunable. \hphantom{} \\

We begin with formulating the above-mentioned setup. For each SO wire, generation of SOC was demonstrated in the $\rm{Rb}^{87}$ atoms which has a ground state with total angular momentum $F=1$, and $m_F=1,0,-1$ multiplets. Lets us assume that the ultracold $\rm{Rb}^{87}$ atoms are optically trapped along the $x$-direction with inter-atomic distance being $a$. Each $\rm{Rb}^{87}$ atom is further regulated with two Raman lasers pointed in the $ \hat{x}+\hat{y}$ and $\hat{x}-\hat{y}$ directions, with slightly detuned frequencies by $\triangle\omega_{\rm L}$. The two electric fields are ${\bf E}_{1}=\sin(k_{\rm L}x)(\hat{x}+\hat{y})$, and ${\bf E}_{2}=\sin(k_{\rm L}x+\Delta\omega_{\rm L} t)(\hat{x}-\hat{y})$, where $k_{\rm L}$ is the wavevector of the lasers. For this setup, the lowest order light-matter interaction term extends upto the rank-1 (vector) terms, giving rise to light shift interaction with atoms as:
\begin{equation}
H=\Omega^{(0)}{\bf E}_{1}\cdot {\bf E}_{2}+\Omega^{(1)}{\bf E}_{1}\times {\bf E}_{2}\cdot \textbf{F},
\end{equation}
where $\Omega^{(i)}$ are the corresponding interaction strengths. Since ${\bf E}_{1}$ and ${\bf E}_{2}$ are orthogonal to each other, first term disappears (henceforth we drop the superscript in $\Omega^{(1)}$). In the second term, ${\bf E}_{1}\times {\bf E}_{2}$ appears as a magnetic field to the atoms and couples to the total angular momentum ${\bf F}$ of the atom (nucleus' total moment + outermost electron's spin momentum) thereby producing SOC. It is observed that the $|m_F=+1\rangle$ state in $\rm{Rb}^{87}$ atom lies at a much higher energy than the other two multiplets\cite{lin} and thus can be neglected. Following adiabatic elimination method (see supplementary materials), we can remove the $|m_F=+1\rangle$ state and obtain an effective 2-level model involving $m_{F}=0$ and $m_{F}=-1$ pseudospin states, defined as the $|\uparrow \rangle$ and $|\downarrow \rangle$ basis, as:\cite{lin}
\begin{equation} \label{eq: 2}
 H_{\rm A}=\xi_k\ \mathbb{I}_2 +\frac{\Omega}{2}\sigma_z + \alpha_R k_x\sigma_y.
\end{equation}
Here, the non-interacting dispersion for each SO wire is $\xi_{k}$. $\sigma_i$ are the usual $2\times 2$ Pauli matrices and $\mathbb{I}$ is the identity matrix. $\alpha_R=\hbar^2k_{\rm L}/m^*$ is the SOC strength, which is tunable both externally (by laser wavelength) and internally (by atom's effective mass $m^*$). 

The SOC is reversed in the adjacent `B' SO wire by reversing the ${\bf E}_{1}$ laser, while keeping ${\bf E}_{2}$ laser the same,\cite{footnote_setup} as shown Fig.~\ref{fig1}(a). This reverses the SOC as well as the intrinsic Zeeman like term to $-\Omega$ in the `B' wire, resulting in the corresponding Hamiltonian as: $H_{\rm B}({\bf k},\Omega)=H_{\rm A}(-{\bf k},-\Omega)$ .

We find that for topological reasons, the $k$-dependent tunneling matrix element between `A' and `B' wires should carry a phase\cite{footnote_complexhopping}. Given the flexibility of the optical lattice, this can be achieved in multiple ways. A simple possible method would be to allow staggered hoppings between the upper- and lower-nearest neighbor wires, as in the case of the Su-Schrieffer-Heeger (SSH) lattice.\cite{ssh1} This is modeled by different hopping parameters from `B' to the top  ($t$) and bottom ($t^{\prime}$) `A' wires as shown in Fig.~\ref{fig1}(b). In the corresponding $k$-space, this leads to the net inter-wire hopping:  $T(k)=-te^{ik_yb}-t^{\prime}e^{-ik_yb}$, where $b$ is the inter-wire distance. 

For analytical solutions of the bulk energy states and the $Z_2$ invariant, we expand the Hamiltonian in the basis of Dirac matrices. The calculation becomes simpler if we set $t^{\prime}<<t$, and without any loss of generality, we set $t^{\prime}=0$. This does not  change the overall band topology and the $Z_2$ invariant as subsequently confirmed with numerical simulation by inserting back the finite $t^{\prime}$ term. For the lattice generalization, we assume a nearest neighbor hopping for both dispersions and SOC which yield $\xi_k= -\gamma_1\cos{(k_xa)}-\gamma_2\cos{(k_yb)}-\mu$, and $\alpha_{k}=-i\alpha_R \sin{(k_xa)}$, where $\gamma_{1,2}$ are the usual tight-binding parameters between same spin species in the $x$ and $y$ directions, respectively, $\mu$ is the chemical potential, and $a$ and $b$ are the corresponding inter-atomic distances. Therefore, we can express the total Hamiltonian in a four-component spinor defined as $(|{\rm A}_\uparrow \rangle$, $|{\rm A}_\downarrow \rangle$, $|{\rm B}_\uparrow\rangle$, $|{\rm B}_\downarrow\rangle)$ (where `A' and `B' stand for atoms on `A' and `B' SO wires, respectively):
\begin{eqnarray}
H(k)=\left(
 \begin{array}{ c c c c }
\xi_k+\frac{\Omega}{2} & \alpha_k        &  -te^{ik_yb}                      & 0\\
-\alpha_k   & \xi_k-\frac{\Omega} {2}  & 0 & -te^{ik_yb}                    \\
-te^{-ik_yb}                  &                  0 & \xi_k-\frac{\Omega}{2} & -\alpha_k\\
 0                  &                       -te^{-ik_yb}   & \alpha_k  &\xi_k+\frac{\Omega}{2}
\end{array} \right).
\label{Hamp}
\end{eqnarray}
The corresponding eigenvalues are $E_{\pm}(k)=\xi_k\pm \frac{1}{2}\sqrt{4|\alpha_k|^2 + \Omega^2 + 4 t^{2}}$. The TR operator for the above spinor can be defined as $\tau=-i \mathbb{I}\otimes\sigma_y\mathbb{K}$\footnote{Interchanging the basis to $\{A_\uparrow,B_\downarrow,B_\uparrow,A_\downarrow\}$ retrieves the familiar TR operator $\tau^\prime=-i \sigma_x\otimes \sigma_y$, $H\rightarrow H^\prime$ and hence $\tau^\prime H^{\prime*}(\textbf{k},\Omega)\tau^{\prime -1}\ =\ H^\prime (-\textbf{k},-\Omega)$.}, where $\mathbb{I}$ is a 2$\times$2 identity matrix, and $\mathbb{K}$ is the complex conjugate. The system is TR invariant as $\tau H^*(\textbf{k}, \Omega) \tau^{-1}\ =\ H(-\textbf{k},-\Omega)$, where $\Omega\rightarrow -\Omega$ under TR operation since it represents spin splitting. Therefore, the full Hamiltonian is TR invariant despite it is broken in each $H_{\rm A,B}$ block.  We have subsequently calculated the spin expectation values and shown that the total magnetic moment always vanishes in our setup, further supporting the TR invariance of the Hamiltonian (see Supplementary Material). 

\begin{figure}[t]
\centering
\rotatebox[origin=c]{0}{\includegraphics[width=0.99\columnwidth]{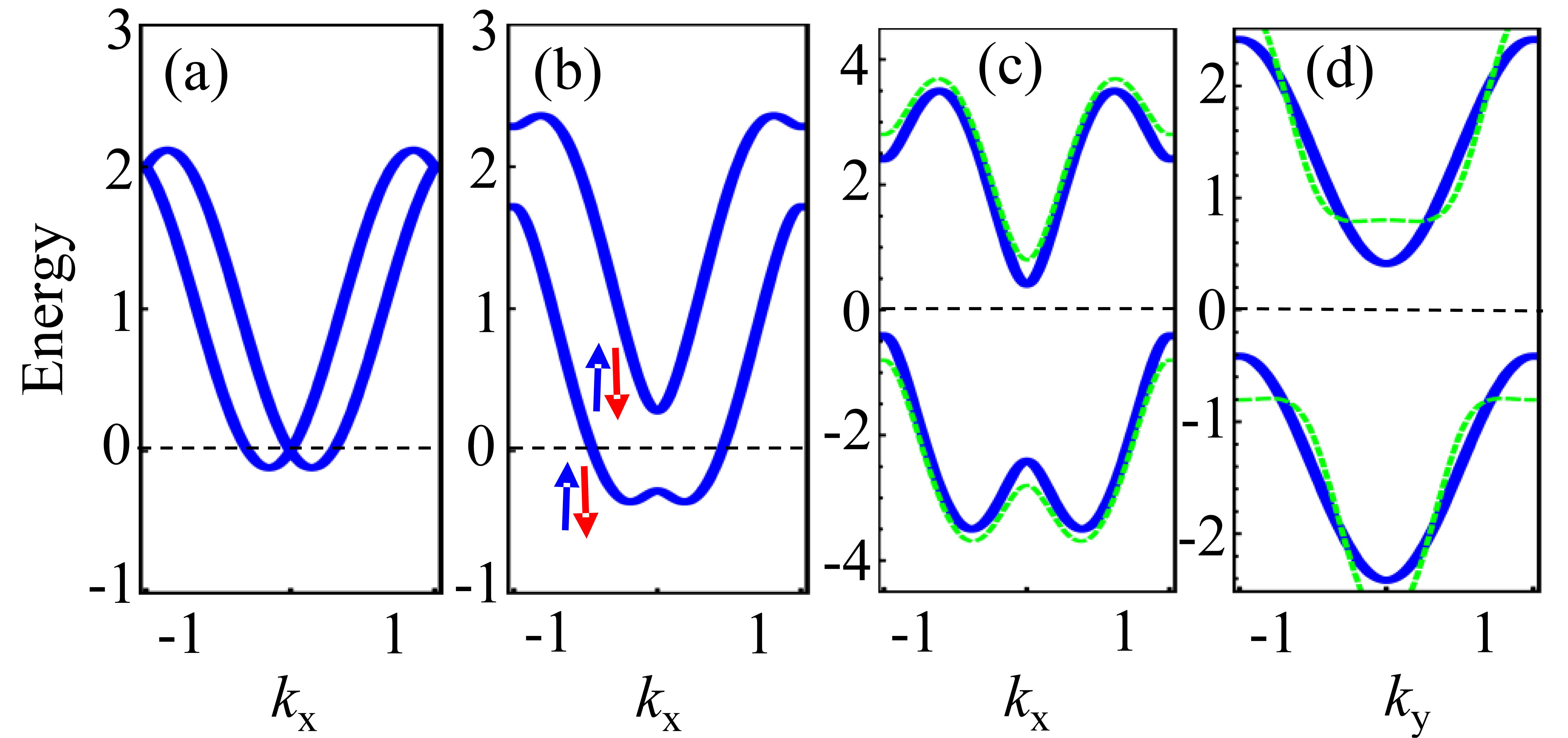}}
\caption{Band evolution towards the formation of TI. All the quantities are measured in units $\gamma_{1,2}= 1$. (a) Band structure of a single SO wire with $\Omega=0$, exhibiting Kramer's degeneracy at the $\Gamma$ point. (b) Calculated band structure of two coupled SO wires, with $\pm\alpha_k$ and $\pm\Omega$. The  gap at the $\Gamma$ without breaking the TR symmetry (each state here is spin-degenerate) is obtained for $|\Omega|=0.2$, $t=0.2$, and $\alpha_R<<\gamma_1$. This situation is the same as illustrated in Fig.~\ref{fig1}(d). (c) As SOC strength is increased above $\alpha_R>>\gamma_1$, the valence band is fully inverted at all $k_x$ points, indicating the emergence of an insulating state. (d)  A representative band structure along the $k_y$-direction. In (c,d), the parameters used are $\Omega=t=1$, and $\alpha_R=3$. Green dashed line shows the same band with including $t^{\prime}=0.5$.
}\label{fig2}
\end{figure}

The TR invariance makes each band doubly degenerate, while the $\Gamma$-point is four-fold degenerate (two spins and two valleys) in the absence of $\Omega$ and $t$; see Fig.~\ref{fig2} and also the figure the supplementary material. Therefore, without breaking the TR symmetry, the $\Gamma$ point can be gapped out with a finite value of $\Omega$ [see Fig.~\ref{fig2}(b)]. The valence band is gradually inverted at all $k$-points as SOC $\alpha_R$, and inter-wire hopping $t$ are increased above their corresponding kinetic energies (i.e. $\gamma_{1,2}$), see Fig.~\ref{fig2}(c-d). As $t$ becomes comparable to the intra-wire kinetic energy, inter-wire hopping becomes more favorable. As a spin-up atom hops from one wire to the next one, its propagation direction becomes reversed, due to opposite SOC. Finally, by hopping back to the first wire, it forms a `chiral orbit' (opposite chirality for the spin-down atom) in the bulk, with an associated $Z_2$ topological invariant.

Unlike the typical Hamiltonians of QSH insulators\cite{bhz,kane}, where two $2\times 2$ blocks for different spins are decoupled, in our case they are coupled by the SOC, mixing the spin states. Therefore, a simple Chern number for each spin cannot be deduced. Here the topological invariant can be calculated from the TR operator\cite{kane,fu,soluyanov}, or from the Wilson loop.\cite{yu} We discuss the former procedure here. According to the Kane-Mele formalism,\cite{kane} the $Z_2$ invariance can be calculated by counting the number of pairs of zeros in the Pfaffian of the overlap matrix defined as:  $P({\bf k})={\rm Pf}[\langle u_{i}({\bf k})|\tau|u_{j}({\bf k})\rangle]$, and $|u_i({\bf k})\rangle$ is the Bloch state for the $i^{\rm th}$-filled band. For the Hamiltonian in Eq.~\ref{Hamp}, the valence band is twofold degenerate, so the Pfaffian is just the $i\ne j$ component of the overlap matrix. The exact evaluation of $P({\bf k})$ comes out to be
\begin{equation}
P(k) = 2\ (1 + e^{-2 i k_yb}).
\label{Eq:Pk}
\end{equation}
It is interesting to note that $P(k)$ only depends on the phase associated with the inter-wire hopping, and is parameter free. This justifies the inclusion of staggered inter-wire hopping allowing the survival of this complex momentum dependent phase. The loci of the nodes in $|P(k)|$ is $k_y^*=\pm\frac{\pi}{b}(n + \frac{1}{2})$ ($n$ is integer), for any value of $k_x$, except at the TR invariant point. Each node at $+k_y^*$ renders a positive winding number, also referred by `vorticity' or `chiral orbit', while that at $-k_y^*$ yields a negative winding number since $\pm k_y^*$ are related by TR invariance. The $Z_2$ invariance can therefore be evaluated by the winding number of $P(k)$ as \cite{kane,fu}:
\begin{equation}
\nu =\ \frac{1}{2 \pi i} \oint_{C_{1/2}} d \textbf{k}. \nabla_{\textbf{k}}\big( \log{[P(\textbf{k})+ i \delta]} \big)\ \rm{mod}\ 2,
\label{Eq:TI}
\end{equation}
where $C_{1/2} $ denotes that the integral is over half of the BZ, $ k \in [0, \frac{\pi}{b}]$, enclosing either $k_y^*$ or $-k_y^*$ point. As the contour $C_{1/2}$ encloses a single Pfaffian node, the integral gives $\nu=1$, indicating the presence of non-trivial bulk topology with odd pair of edge states. To establish the robustness of the topological invariant of this setup, we have also evaluated the $Z_2$ invariant $\nu$ by inserting back the $t^{\prime}$ term in the Hamiltonian. We find that the inclusion of  the $t^{\prime}$ term keeps the Pfaffian unchanged and we still obtain $\nu =1$ as long as $t^{\prime} \neq t$.  We emphasize that the non-trivial $Z_2$ invariance is ensured by the geometry of our setup in which the adjacent `A' and `B' wires serve as TR conjugate to each other which enables odd number of TR partner inversion in half of the BZ, and therefore each half encloses a single node of $P(k)$.

\begin{figure}
\centering
\rotatebox[origin=c]{0}{\includegraphics[width= 0.99\columnwidth]{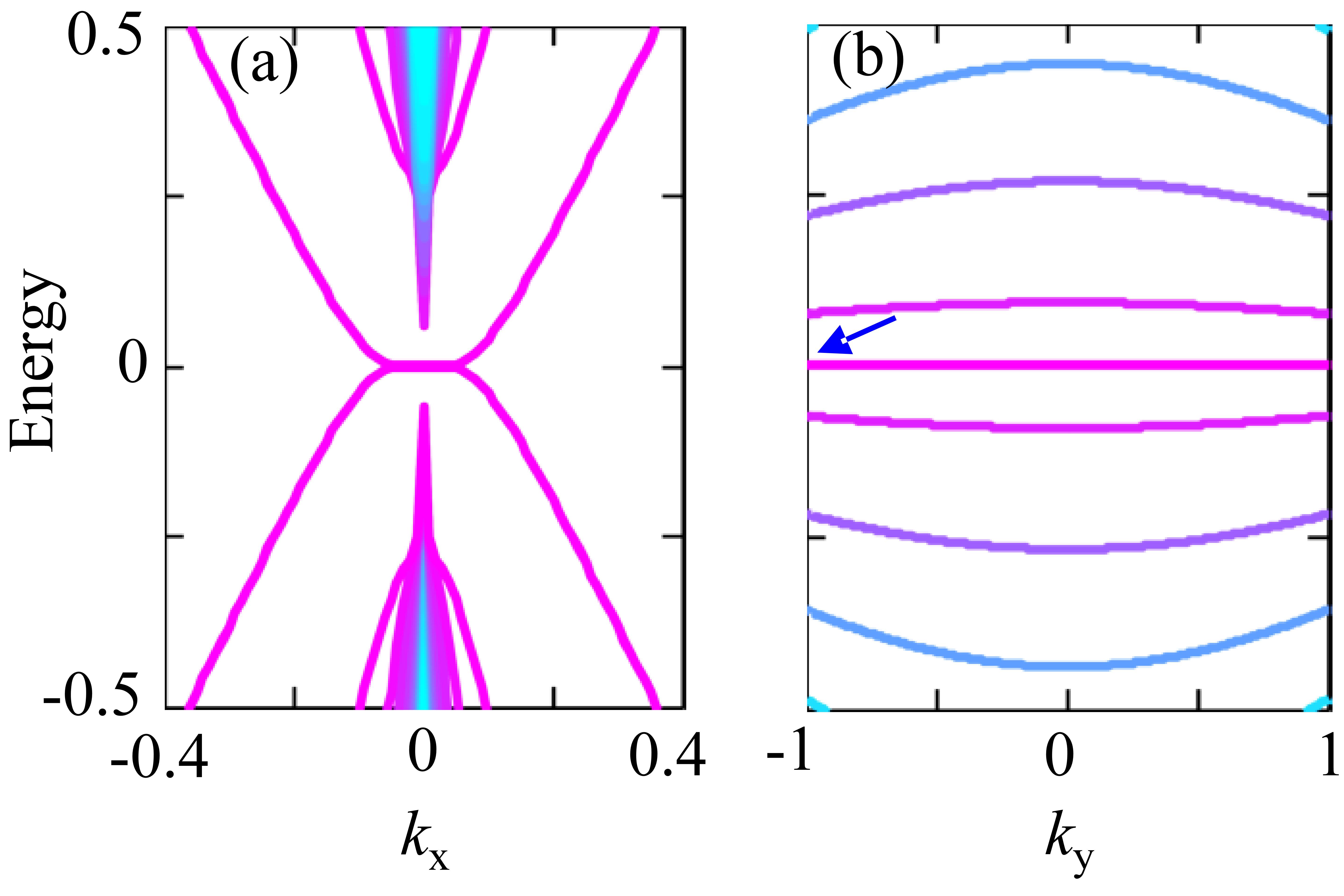}}
\caption{(a) Quantum well and edge states for the supercell calculation with open boundary condition along the $y$-axis and periodic lattice along the $x$-axis. Cyan and magenta color represents that the corresponding state originates from the bulk to the edge, respectively. As discussed in the main text, we obtain gapless edge state with quadratic dispersion owing to the Mirror symmetry. We set $t^{\prime}= 0.1 t$, $t = 0.1 |\gamma_{1}|$ and $\alpha_R \sim 10 |\gamma_{1}|$. (b) Evolution of the edge states along the perpendicular direction to the configuration in (a). As in the case of SSH lattice, a gapless bound state (blue arrow) arises for $t^{\prime}>>t$, and SOC $\alpha_R << t^{\prime}$. As SOC is turned on, these states split. The parameter values for this plot are $\alpha_R = |\gamma_{1,2}|=1, t=1$ and $t^{\prime} = 10 t$.}\label{fig3}
\end{figure}

The bulk-boundary correspondence of the TI dictates the presence of the conducting edge states, connecting the bulk conduction and valence bands.\cite{TIreviewTD,TIreviewCK,TIreviewSCZ} The properties of the edge states for the present setup are studied both analytically and numerically using the bulk Hamiltonian from Eq.~(\ref{Hamp}). For the edge states, the crystal symmetries play important roles. Terms involving $k_x$ and $k_y$ variables are decoupled into different off-diagonal terms in the Hamiltonian (Eq.~\ref{Hamp}) with different coupling constants, leading to a rotational ($C_4$) symmetry breaking. This makes the two edge states behave differently. As a byproduct, the system possess Mirror symmetry in the $x$-direction, which restricts that the eigenvalues should be even in $k_x$. Therefore, the leading term in the corresponding edge state becomes quadratic in momentum, rather than a quintessential linear dispersion. However, it remains helical owing to the SOC. 

The edge parallel to the $x$-axis is made of just a single SO wire lying at $y=0$, coupled to the non-trivial bulk for $y>0$ and the vacuum at $y<0$. To make the analytical calculation manageable, we solve the  bulk Hamiltonian in Eq.~\eqref{Hamp} with periodic boundary condition along the $x$-direction, but open boundary condition along the $y$-direction with two SO wires. We solve the Schr\"odinger equation in the continuum limit ($\alpha_k\rightarrow i\alpha_R k_x$), obtain the helical low-energy edge states (up to the quartic term) as  
\begin{equation}
E_y (k_x)\ =\ \pm \bigg(t + \frac{\alpha_R^2}{2 t}\ k_x^2 -\frac{\alpha_R^4 }{8 t^3}\ k_x^4\bigg).
\label{edge}
\end{equation}
We note that the two edge states are `apparently' gapped by $t$, and the gap vanishes at some finite value of the momentum for a given parameter set ($k_x \approx \pm 0.015$ for the present case, where $t=0.1$ and $\alpha_R=15$).  This is due to the finite size effect of the geometry. As the number of wire is increased, the gap decreases gradually and eventually vanishes at $k_x=0$. This result is confirmed by numerical simulation of a system with 50 pairs of SO wires, and the corresponding results are shown in Fig.~\ref{fig3}. The Fermi velocity of the edge state is proportional to the tunable parameters SOC strength $\alpha_R$, and the hopping amplitude $t$, see Eq.~\eqref{edge}. We have also estimated the decay length scale of the edge state into the bulk, and find that it is directly proportional to $\Omega$, and thus can be monitored externally (see Supplementary Materials, section C. for further details).

For our choice of the staggered hopping along the $y$-direction, the corresponding edge state properties follow that of the SSH model. In the limit of $t^{\prime}\rightarrow 0$, the edge consists of a dimer of two atoms coming from the `A' and `B' wires. The dimer remains decoupled from the bulk, and therefore the edge state is gapped. As $t^{\prime}$ is increased, the corresponding gap decreases and gapless end states arise in the limit of $t^{\prime}>>t$ and $\alpha_R\rightarrow 0$. The present setup, however, makes the end states helical as SOC is turned on. In summary, for our proposed setup, conducting and helical edge states arise along the edges parallel to the direction of SOC wires, while the dispersionless end modes along the perpendicular direction can be tuned from gapped to gapless with SOC and inter-wire hopping. Therefore, these edge states promise multifunctional applications: while the conducting edge states can be used for transport related applications, the other edge can be used for optical switch and transistor related applications owing to the tunable bang gap.   

\begin{figure}
\centering
\rotatebox[origin=c]{0}{\includegraphics[width= 0.99\columnwidth]{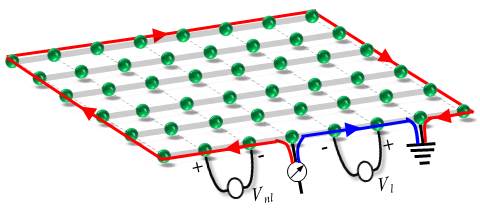}}
\caption{A non-local measurement geometry to detect the chiral edge states of the proposed TI. The expected local voltage ($V_l$) drop arises between the source and the ground, while an additional non-local voltage ($V_{nl}$) is expected to arise due to the chiral nature of the edge states for the $Z_2$ TI. The right-handed and left-handed arrows attached to the corresponding lines dictate the local and non-local currents, respectively.}\label{fig5}
\end{figure}

In Fig.~\ref{fig5} we provide the relevant setup to detect the chiral edge states for the present case.\cite{Roth} We attach a current source and a ground to two different edge atoms. The voltage drop occurring between them is the expected local voltage ($V_l$). An additional current flow is expected to arise in the other direction (reaching to the ground from the other side of the edge), dictated by left moving arrow in Fig.~\ref{fig5}. This occurs due to the chiral property of the edge state in $Z_2$ TI. Measuring the corresponding non-local voltage drop ($V_{nl}$) with opposite sign gives us a concrete prove to the emergence of the TI in the bulk. The edge current is expected not to dissipate, in principle, with distance from the source, due to topological protection. This can be checked by measuring voltage $V_{nl}$ farther from the source atoms. The absence of bulk conductivity can be easily verified here by measuring the voltage drop in the inner atoms.

In condensed matter systems, quantum SO wires are routinely grown in Bi-, Pb- and related elements based atomic wires with tunable SOC.\cite{QW1,QW2} Also as mentioned earlier, for the cases where the bulk (Dresselhaus) and surface (Rashba) SOC in non-centrosymmetric materials cancel each other, a similar 1D SOC arises. For such systems, the generation of tunable TI follows similarly. Generalization to a 3D TI can be obtained easily. We can seek to build a `strong' 3D TI starting with the above-obtained 2D setup. For the second layer, we need to place another 2D layer, with `B' wire sitting on top of the `A' wire and vice versa. That means for each `A' wire the surrounding nearest-neighbor lattice wire along all directions should be the `B' wire with opposite SOC. Such a setup will lead to a TI with odd number of band inversion (at the $\Gamma$-point only) in the entire BZ, a criterion for the strong TI.\cite{TIreviewTD,NCTD} Given that the engineered TIs in optical lattice harness remarkable flexibility and tunability, both bulk and boundary states can be easily manipulated. This can provide a versatile substitute for the topological solid state materials, seeking to overcome of materials challenges embedded in the present materials selections.

{\bf Acknowledgments}
We thank Aveek Bid for suggesting the experimental setup. We thank Srinivas Raghu, Benjamin Lev, Monika Schleier- Smith, Diptiman Sen and Jainendra Jain for useful discussions. SR would also like to thank CSIR for financial support during the work. The work is facilitated by the computer cluster facility at the Department of Physics at the Indian Institute of Science.



\section*{Supplemental Material}
\subsection{Setup and Hamiltonian}
The model setup for the generation of 1D SOC in optical lattice is given in Ref.~\cite{lin} and the corresponding effective Hamiltonian is obtained in the same paper. Here we deduce the full Hamiltonian with the microscopic details, and show that the generation of their bilayer extension for our purpose is viable. The basic setup consists of a periodic array of ultracold bosons designed in optical lattice. We consider that the ultracold atoms are in the ground state with total angular momentum $F=1$, with three states $m_F$= +1, 0 and -1, which as denoted by $|+1\rangle$, $|0\rangle$, and $|-1\rangle$.  $^{87}$Rb atom is such a bosonic atom with $F=1$ ground state which was utilized for the SOC generation\cite{lin}. We start with a three component spinor for the three states, and derive the Hamiltonian in the rank$-3$ Pauli matrices are labeled as $\sigma_{3,i}$:
$$
\begin{pmatrix}
0&1&0\\
1&0&1\\
0&1&0\\
\end{pmatrix},
\begin{pmatrix}
0& - i &0\\
i &0& - i\\
0& i &0\\
\end{pmatrix},
\begin{pmatrix}
1&0&0\\
0&0&0\\
0&0&-1\\
\end{pmatrix},
$$
for $i=x,y$, and $z$, respectively. Finally, we will find in the following sections that the $|+1\rangle$ is pushed significantly up in energy compared to the other two states, allowing very weak hybridization with the other states. Therefore, we can write the Hamiltonian as an effective 2-level system. 
 
An 1D SOC is created (say along $x$-direction) when two Raman lasers are aligned orthogonal to each other, along $\hat x\pm\hat{y}$ directions, and are detuned from each other by $\Delta\omega_{\rm L}$, respectively. In this case, the light-matter interaction basically has three terms: Interaction between the electric fields of the lasers and angular momentum of the Rb-atom, and two Zeeman-like (paramagnetic and diamagnetic) terms arising from the bias magnetic field coupled to the spin. We denote them as $H_{\rm R}$,  $H_{\rm P}$, and $H_{\rm D}$. Considering also a generalized kinetic energy term of the bosons in a rectangular optical lattice as discussed in the main text, denoted by $H_{\rm K}$, we obtain the full Hamiltonian as
\begin{equation}
H= H_{\rm K}  + H_{\rm P} + H_{\rm D}+ H_{\rm R}.
\label{Ham}
\end{equation}
The last term is crucial for the SOC. The total angular momentum ${\bf F}$ has contributions from both the nucleus and the last orbital electron. If the angular momentum is $\textbf{I}$ and the nuclear magnetic moment is $\textbf{J}$, then the total angular momentum is $\textbf{F=I+J}$. The atom-light shift interaction with two monochromatic light fields (${\bf E}_{1,2}$) detuned off resonance, have a generic Hamiltonian,
\begin{equation}
H_{\rm R }=\alpha_{ij}E_{1 i} E_{2 j} \,.
\end{equation}
Here $\alpha_{ij}$ denotes the rank of the interaction, i.e, $\alpha_{ij}=\delta_{ij}$ for scalar (rank 0) and $\alpha_{ij}=\epsilon_{ijl}F_l$ for vector (rank 1) component and so on. Restricting the Hamiltonian upto the rank-1 (vector) light shift interaction with the spin states of the atom, we get 
\be
H_{\rm R} =\Omega^{(0)}{\bf E}_{1}\cdot {\bf E}_{2}+ \Omega^{(1)}{\bf E}_{1}\times {\bf E}_{2}\cdot \textbf{F}.
\label{HR1}
\ee
The strength of each interaction is denoted by $\Omega^{(i)}$. Since the two lasers are aligned orthogonal to each other and intersect at the origin, the first term in Eq.~(\ref{HR1}) vanishes. Henceforth, we drop the superscript `(1)' from $\Omega^{(1)}$. The rank-1 light interaction acts as an anisotropic magnetic field which couples to the atom's total angular momentum. We assume that both the laser wavelengths are denoted by $k_{\rm L}$, with frequencies $\omega_{\rm L}$, and $\omega_{\rm L}+\D\omega_{\rm L}$. We set $\omega_{\rm L}=0$ (which eventually drops out otherwise), we get 
\begin{align}
H_{\rm R}=&\Omega [\sin (k_L x)(\hat{x}-\hat{y})\times\sin (k_L x+\D\omega_L t)(\hat{x}+\hat{y})]\cdot \textbf{F}\nonumber\\
=& 2\Omega [\cos(2k_L x+\D\w_L t)-\cos (\D\w_L t)]\s_{3,z}\,.
\end{align}
Furthermore, since the second term in the above equation gives only a constant energy shift to the ground state energy, we can also drop this term, yielding $H_{\rm R}=\Omega \cos[2k_L x+\D\w_L t] \s_{3,z}$. 

The remaining terms in the Hamiltonian are usual. In the continuum limit, we take anisotropic effective mass for the intra-wire and inter-wire hoppings, which gives 
\be
H_{\rm K}=\xi_{k}\mathbb{I}_3= \left[E_{\rm L}+\frac{\hbar^2}{2}\left(\frac{k_x^2}{m_1}+\frac{k_y^2}{m_2}\right)\right]\mathbb{I}_3,
\ee
where $\mathbb{I}_3$ is the $3\times3$ identity matrix. $\xi_k$ is the non-interacting dispersion term for each spin states. $E_L$ is the energy shift due to the lasers. $\vec{B}\cdot \vec{\mu}$, is the paramagnetic term coming from the externally applied bias magnetic field $\vec{B}$. We choose the bias magnetic field  along the $y$-direction which gives $H_{\rm P}=\delta\s_{3,y}$, where $\d=-\m_B B_y$. The diamagnetic term $H_{D}$ gives a quadratic Zeeman effect, denoted by $ \hbar \omega_q$, shifting the the  $|+1\rangle$ state further from the $|0 \rangle$ and $|-1\rangle$. 

With all the terms discussed above, we can now write down the 3-level Hamiltonian in the basis $|+1\rangle,|0\rangle$ and $|-1\rangle$ as follows,
\beq
\begin{split}
H=&\xi_{k}\mathbb{I}_3+ \begin{pmatrix}
\hbar\w_q&0&0\\
0&0&0\\
0&0&0\\
\end{pmatrix}\\
&+\d\s_{3,y} +\Omega \cos[2k_L \hat{x}+\D\w_L t]\s_{3,z}.
\end{split}
\eeq
Since we choose the $\hat{y}$ axis as the natural quantization axis, we can perform a global rotation as $\s_{3,y}\rightarrow \s_{3,z},\s_{3,x}\rightarrow \s_{3,y}$ and $\s_{3,z}\rightarrow \s_{3,x}$.
Then we add a constant term $ \frac{\d}{2}\ \mathbb{I}_3$ which shifts each of the levels by $\frac{\d}{2}$. Thereby second and third terms can now be combined to obtain 
\hphantom \\
\be\label{Hm1}
\begin{split}
H=&\xi_{k}\mathbb{I}_3+ \begin{pmatrix}
\frac{3\d}{2}+\hbar\w_{q}&0&0\\
0&\frac{\d}{2}&0\\
0&0&-\frac{\d}{2}\\
\end{pmatrix} \\
& +\Omega \cos[2k_{\rm L} x+\D\w_{\rm L} t]\s_{3,x}.
\end{split}
\ee
$|+1\rangle$ state is thus separated by a relative large energy scale $\hbar(\delta+\w_q)$ from $|0\rangle$ and $|-1\rangle$ states. Next we perform another a rotation about the $\hat{z}$ axis to go to the co-moving frame rotating with frequency $\D\w_{\rm L}$. This helps eliminate the $\w_{\rm L}$ term from the Hamiltonian without changing anything else in it. This is done by using the so-called Rotating Wave Approximation (RWA).\cite{fujii,sanchez} The RWA procedure is analogous to going from the Schrodinger picture (where the states are time evolving) to the Heisenberg picture (where the operators are time evolving and the states are not). In this case we want to go to a frame where  this is static. A familiar choice of the transformation matrix is given by,
\be
U=\exp[-i (\D\w_L t/2)\s_{3,z}]\,.
\label{Op}
\ee
The Hamiltonian thus transform to $H'=U H U^\dagger\,$. The identity matrix remains invariant under this rotation. Therefore, the first and the second term, which can be decomposed into terms contained $\mathbb{I}$ and $\s_{3,z}$), also remain the unchanged under this $U$ transformation.  
$U$ gives a non-trivial effect for the SO term which can be seen as follows: 
\begin{eqnarray}
&&U \cos[2k_{\rm L} x+\D\w_{\rm L} t]\s_{3,x} U^\dagger \nonumber \\
&&\qquad\qquad\qquad=\cos(2k_{\rm L} x)U\cos(\D\w_{\rm L} t)\s_{3,x}U^\dagger\nonumber\\
&&\qquad\qquad\qquad~+\sin(2k_{\rm L} x)U\sin(\D\w_{\rm L} t)\s_{3,x}U^\dagger \,,\nonumber\\
\end{eqnarray}
Let us first write down the $\s_{3,i}$ operators in a convenient fashion as, $\s_{3,\pm}=(\s_{3,x}\pm\s_{3,y})/2$ and 
further $e^{i\D\w_L t}=\cos \D\w_L t+i \sin \D\w_L t$. Thus we can write 
\begin{align}
\cos \D\w_L t \s_{3,x}&=\frac{1}{2}(e^{i\D\w_L t}+e^{-i\D\w_L t})(\s_{3,+}+\s_{3,-})\nonumber\\
&=\frac{1}{2}(e^{i\D\w_L t}\s_{3,+}+e^{-i\D\w_L t}\s_{3,-}\nonumber\\
& ~~~~+e^{i\D\w_L t}\s_{3,-}+e^{-i\D\w_L t}\s_{3,+})\,.
\end{align}
Using the facts that,
$U \s_{3,\pm}U^\dagger=e^{\mp i\D\w_L t}\s_{3,\pm}\,$ 
we can show that the first term becomes
\be
\begin{split}
U \cos(\D\w_{\rm L} t)\s_{3,x}U^\dagger=\frac{1}{2}&(\s_{3,+}+\s_{3,-}+e^{2i\D\w_{\rm L} t}\s_{3,+}\\
&+e^{-2i\D\w_{\rm L} t}\s_{3,-})\,.
\end{split}
\ee
Neglecting the terms proportional to $e^{\pm i\D\w_L t}$, since they represent rapidly oscillating terms, we get,
\be
U \cos(\D\w_{\rm L} t)\s_{3,x}U^\dagger=\frac{1}{2}\s_{3,x}\,.
\ee
Similarly for the other term we get,
\be
U \sin (\D\w_{\rm L} t) \s_{3,x}U^\dagger= -\frac{1}{2}\s_{3,y}\,.
\ee
We thus get the resultant Hamiltonian in the rotated frame as,
\be\label{3lh}
\begin{split}
H=&\xi_{k}\mathbb{I}_3+\begin{pmatrix}
\frac{3\d}{2}+\hbar\w_q&0&0\\
0&\frac{\d}{2}&0\\
0&0&-\frac{\d}{2}\,.
\end{pmatrix} \\
&+\frac{\Omega}{2}\left[\cos(2k_{\rm L}x)\s_{3,x}-\sin(2k_{\rm L} x)\s_{3,y}\right]\,.
\end{split}
\ee

\noindent
{\bf Effective two levels Hamiltonian.} As discussed before, the $|1\rangle$ is shifted to much higher in energy compared to other two states due to the diamagnetic term. Therefore, we can neglect this terms and obtain an effective $2\times 2$ Hamiltonian described by usual Pauli matrices $\sigma_i$ [The full derivation of $2\times2$ Hamiltonian using Adiabatic Elimination method is given in Sec.~\ref{Sec:AEM}]. The corresponding Hamiltonian takes the form 
\beq\label{h2}
H=\xi_{k}\mathbb{I}_2+\frac{\d}{2}\s_z 
 +\frac{\Omega}{2}\left[\cos(2k_L\hat{x})\s_x
-\sin(2k_L\hat{x})\s_y\right].\nonumber\\
\eeq

Now apply a pseudo spin rotation about the $\hat{z}$ axis with a phase of $\theta(x)=2k_{\rm L} x$, which gives $U_x=\exp[i k_{\rm L} x \s_z]\,$. As done previously, we find that,
\be
U_x\left[\cos (2k_L \hat{x})\s_x-\sin (2k_L \hat{x})\s_y\right]U^\dagger=\s_x\,.
\ee
There is, however, an important different in this rotation compared to the rotation for operator in Eq.~(\ref{Op}). Here $U_x$ does not commute with the kinetic energy term $H_{\rm K}=\xi_k\mathbb{I}$. This is because the leading term in $\xi_k$ is $k_x^2$, which gives rise to an non trivial term linear in $k_x$ in the SOC Hamiltonian under $U_x$ rotation. This can be seen explicitly as
\be
U\xi_{k}U^\dagger= E_{\rm L} +  \frac{\hbar^2}{2m_1} Uk_x^2U^{\dagger}+\frac{\hbar^2}{2m_2}k_y^2\\,.
\ee
Terms containing $k_y$
do not change under $U_x$ rotation. We take an infinitesimal representation of the operator $U$ as,
\be
U=1+i\frac{\theta}{2}\s_z\,.
\ee
Thus with $\theta=2k_L \hat{x}$ we have,
\begin{align}\label{kxtrans}
\frac{\hbar^2}{2m_1}e^{i\theta/2 \s_z}k_x^2e^{-i \theta/2 \s_z}&=\frac{\hbar^2}{2m_1}(1+i\frac{\theta}{2}\s_z)k_x^2(1-i\frac{\theta}{2}\s_z)\nonumber\\
&=\frac{\hbar^2k_x^2}{2m_1}+\frac{i\hbar^2}{4m_1}[\theta, k_x^2]\s_z\nonumber\\
&=\frac{\hbar^2k_x^2}{2m_1}+\frac{\hbar^2k_Lk_x}{m_1}\s_z\,.
\end{align}
Thus the total 2-level Hamiltonian after the rotation becomes,
\be
H=\xi_{k}\mathbb{I}_2+\frac{\d}{2}\s_z+\frac{\Omega}{2}\s_x+\frac{\hbar^2k_Lk_x}{m}\s_z\,.
\ee
Finally, we employ a global rotation $\s_z\rightarrow\s_y$, $\s_y\rightarrow\s_x$ and 
$\s_x\rightarrow\s_z$, to get the final form of the Hamiltonian as,
\be\label{2lh}
H=\xi_{k}\mathbb{I}_2+\frac{\Omega}{2}\s_z+\frac{\d}{2}\s_y+\frac{\hbar^2 k_L k_x}{m}\s_y\,.
\ee
 
In the final Hamiltonian, we can easily recognize that the last term gives a 1D SOC. We denote the SOC coupling strength as $\alpha_R=2\hbar^2k_{\rm L}/2m_1$. It is interesting to see that the SOC strength is inversely proportional to the band mass, which is opposite to the case for a simple SOC in solid state systems. Therefore, SOC can be tuned here by the effective mass of the electron propagating along the SOC wire, as well as by the wavevector of the incident laser. This constitute the Hamiltonian for the `A' wire in the main text.

Recalling that $\delta$ arises from the external bias, and its value decreases with increasing laser frequency in the actual experiment\cite{lin}, without losing generality, we can set it to be zero. The other term $\Omega$ is proportional to the direction of intrinsic magnetic field of the laser. For our bilayer setup, we set out to obtain an effective Hamiltonian which remains time-reversal invariant. This can be obtained by rotating one of the lasers into the opposite direction in the adjacent wire [see Fig.~1 in the main text]. This has two effects. It creates the SOC in the reverse direction ($-k_x$), as well as change the sign of $\Omega\rightarrow -\Omega$. The explicit form of the Hamiltonian for the `B' wire is then
\be\label{2lh}
H_{\rm B}=\xi_{k}\mathbb{I}_2-\frac{\Omega}{2}\s_z-\alpha_R k_x \s_y\,.
\ee
Thereby we restore the time-reversal symmetry in the total Hamiltonian. We note that due to the presence of the quantum tunneling between the two layers, which is of the form $te^{ik_yb}$, we see that the block Hamiltonians cannot be separated, as was done for the quantum Spin Hamiltonian for the HgTe/CdTe quantum wells in Ref.\cite{bhz}. Therefore, we cannot compute the Chern number for each block. Calculation of the proper $Z_{2}$ invariant from the expectation value of the time-reversal operator,\cite{TIreviewCK,TIreviewSCZ,TIreviewTD,kane} is thus necessary here.   \hphantom{} \\
In the real-space, the hoppings from `B' to the top and the bottom `A' wires are taken to be different ($t\ne t^{\prime}$) without any complex phase associated with it. This staggered hopping naturally leads to a complex Bloch in the momentum space, according to the SSH model (Ref. 23), and serves our purpose. This can be achieved by tuning the distance between the two adjacent wires to be slightly different, as illustrated in Fig.1.(b).  \hphantom{} \\
Two SO wires are not required to be placed on the same plane. If they are placed on different planes (along the $z$-direction) to facilitate preparations, the inter-wire hopping in Hamiltonian (Eq. 3) would still remain the same as long as the distance ($b$) along the $y$ direction is kept to be the same.  

\vspace{0.5cm}

\noindent
{\bf Generalization to Lattice Model.}
The above analysis can be generalized to a lattice model in which we replace $k_x\rightarrow \sin{(k_xa)}/a$, and $k_x^2\rightarrow 2(\cos{(k_xa)}-1)/a^2$, where $a$ is the lattice constant. If we assume the nearest neighbor hopping amplitudes as $\gamma_{1,2}$ along the $x$-, and $y$-directions, respectively, the non-interacting dispersion in 2D becomes $xi_{\vec{k}} = \hbar^2k_x^2/2m_1^*+\hbar^2k_y^2/2m_2^* \rightarrow \gamma_1\cos{k_x} +\gamma_2 \cos{k_y} -\mu$. Here $\gamma_{1}=\hbar^2/m_1^*a^2$, $\gamma_{2}=\hbar^2/m_2^*b^2$, and the chemical potential is $\mu=\hbar^2(1/m_2^*a^2+1/m_2^*b^2)$. 

 The SOC term arises by the same formalism from Eq.~(\ref{kxtrans}) in the following way:
\begin{align}\label{SOClattice}
& \bigg[ e^{i\theta/2 \s_z}(\xi_k)e^{-i \theta/2 \s_z} \bigg]\nonumber\\
& \approx (1+i\frac{\theta}{2}\s_z)(\gamma_1\cos{k_x}+\gamma_2 \cos{k_y}-\mu)(1-i\frac{\theta}{2}\s_z)\nonumber\\
&= \bigg[ (\xi_k)+ \frac{1}{4}[\theta, (\xi_k)]\s_z \bigg]
\end{align}
By expanding $\cos$ term, we perform the commutation operation with $\theta$ with each power of $k_x$ and $k_y$. The commutator of $\theta$ with $k_y$ will naturally give zero. The commutator with $k_x$ will give rise to a $\sin{k_x}$ term, thus an effective 1D SOC, as follows:
\begin{align}
& \bigg[ e^{i\theta/2 \s_z}(\gamma_1\cos{k_x} +\gamma_2 \cos{k_y} -\mu)e^{-i \theta/2 \s_z} \bigg] \nonumber \\
& = \xi_{\vec{k}}- (2\gamma_1 k_L) \sin{k_x} \s_z \nonumber \\
& = \xi_{\vec{k}}- \alpha_ R \sin{k_x} \s_z\,,
\end{align}
where $\alpha_R=2\gamma_1k_L$.

\subsection{Adiabatic Elimination method}\label{Sec:AEM}
Here we elaborate the derivation of the $2\times 2$ Hamiltonian in Eq.~(\ref{h2}) from the $3\times3$ Hamiltonian in Eq.~(\ref{3lh}). Here we use the Adiabatic Elimination method to eliminate the $|+1\rangle$ state. \cite{brion} This method primarily depends on the fact that one of the states (say $|+1\ra$) is so high in the energy compared to the states ($|0\ra$ and $|-1\ra$),  eliminating the $ |+1 \rangle $ state will not alter the band structure . In other words, the time evolution of the excited state is not affected by the ground states. We begin by considering the Hamiltonian in \eqref{3lh} given in the matrix form as,
\be
H=
\begin{pmatrix}
\xi_k+\frac{3\d}{2}+\hbar\w_q&\frac{\Omega}{2}e^{2i k_{\rm L} x}&0\\
\frac{\Omega}{2}e^{-2i k_{\rm L} x}& \xi_k+\frac{\d}{2}&\frac{\Omega}{2}e^{2i k_{\rm L} x}\\\
0&\frac{\Omega}{2}e^{-2i k_{\rm L} x}&\xi_k-\frac{\d}{2}\\
\end{pmatrix}\,.
\ee
Here we have neglected $E_{\rm L}$, the total energy due to the Raman lasers since this only gives a constant energy shift. We take a spinor for the three states $|\pm1\ra$ and $|0\ra$ as
\be
\psi(t)=\begin{pmatrix}
\a(t)\\
\b(t)\\
\g(t)\\
\end{pmatrix}\,.
\ee
where $\a(t)\equiv|+1\ra$, $\b(t)\equiv|0\ra$, and $\g(t)\equiv|-1\ra$, respectively. Substituting $\phi(t)$ in the Schr\''odinger equation $\pd_t\psi(t)=H\psi(t)\,$, gives a set of three coupled differential equations for $\a(t)$, $\b(t)$ and $\g(t)$. We are interested here to study only the effect on $\a(t)$ due to the other states. Thus,
\be
\dot{\a}(t)=\bigg(\xi_k+\frac{3\d}{2}+\w_q\bigg)\a(t)+\frac{1}{2}e^{2i k_{\rm L} x}\Omega \b(t)\,.
\ee
Setting $\dot{\a}(t)=0$ gives,
\be
\a(t)=\frac{\Omega e^{2i k_{\rm L} x}}{2\xi_k+3\d+2\w_q}\b(t)\,.
\ee
Putting this in the remaining set of the coupled differential equations, we immediately see that the spinor containing the two low lying states $\bar{\psi}(t)=(\b(t),\g(t))$ satisfy,
\be
\pd_t\bar{\psi}(t)=H_{2,eff}\bar{\psi}(t)\,,
\ee
where,
\be
H_{\rm 2,eff}=\begin{pmatrix}
\xi_k+\frac{\d}{2}+\frac{\Omega^2}{2\xi_k+3\d+2\w_q}&\frac{\Omega}{2}e^{2i k_{\rm L} x}\\
\frac{\Omega}{2}e^{-2i k_{\rm L} x}&\xi_k-\frac{\d}{2}\\
\end{pmatrix}
\ee
$H_{\rm 2,eff}$ can be put effectively in the basis as
\be
H_{\rm 2,eff}=\xi_k\mathbb{I}_2+\frac{\Omega}{2}\cos(2k_{\rm L} x)\s_x-\frac{\Omega}{2}\sin(2k_{\rm L} x)\s_y+\r \s_z\,,
\ee
where,
\be
\r=\frac{1}{2}\bigg(\d-\frac{\Omega^2}{2\xi_k+3\d+2\w_q}\bigg)\,.
\ee
Expanding $\r$ around $\w_q>\Omega\gg \d$ up to the first sub leading order we get,
\be
\r=\frac{\d}{2}-\frac{\Omega^2}{4\w_q}\,.
\ee
Thus we can see that the coupling $\d$ is effectively modified by the additional term,
\be
\d^{(2)}=-\frac{\Omega^2}{4\w_q}\,.
\ee
Further expanding $\r$ upto one more order, we find that,
\be
\r_1=\r+\frac{2\xi_k+3\d}{8\w_q^2}\Omega^2\,.
\ee
Note that the effective two level Hamiltonian now takes the form,
\be
\begin{split}
H_{\rm 2,eff}=\xi_k\bigg(1+\frac{\Omega^2}{8\w_q^2}\bigg)\mathbb{I}_2 & +\frac{\Omega}{2}\cos(2k_{\rm L} x)\s_x \\
&-\frac{\Omega}{2}\sin(2k_{\rm L} x)\s_y+\r \s_z\,,
\end{split}
\ee
for $k^2\gg \d$. Finally employing the RWA as before, we obtain 
\be
\begin{split}
H'_{\rm 2,eff} &\approx\xi_k\bigg(1+\frac{\Omega^2}{8\w_q^2}\bigg)\mathbb{I}_2  +\frac{\Omega}{2}\cos(2k_{\rm L} x)\s_x \\
& -\frac{\Omega}{2}\sin(2k_{\rm L} x)\s_y+\r\s_z+(\a+\a^{(2)})\s_z,
\end{split}
\ee
where $\a^{(2)}=\frac{\a}{8\w_q^2}\Omega^2\,$, and we can consider $\frac{\Omega^2}{8\w_q^2}\ll1$. Hence we obtain Eq.~(\ref{h2}). 

\subsection{Calculation of edge states}
The edge state calculation follows the same procedure as used earlier,\cite{TIreviewCK,TIreviewSCZ,TIreviewTD,shen} but unlike these models, where the two blocks of the Hamiltonian are decoupled, here they are coupled by either SOC or the inter-wire tunneling. Therefore, the edge state calculation requires special treatment. In our Hamiltonian, the coupling along the $x$- and $y$-directions are different, and the system does not possess the rotational $C_4$ symmetry. Therefore, both edge states have different characteristics, which can be evaluated from the bulk Hamiltonian, owing to the bulk-boundary correspondence of the topological insulator.  Here we discuss the analytical results in the low-energy limits of the edge states. To make the problem manageable with exact diagonalization procedure, we consider the bulk Hamiltonian for two SO wires given in the main text, but relax the periodic boundary condition for the edge under consideration, while keep the periodic boundary condition in the perpendicular direction.. 

\subsubsection{Edge parallel to x-axis}
We first consider the edge parallel to the $x$-axis, or parallel to the SO wire. The edge of this setup is a decoupled SOC wire, lying at, say $y=0$ position. Therefore, the edge state is made of two counter-propagating chiral spin states along the $x$-direction, while decaying exponentially along the $y$-direction. An important symmetry to recognize here is that the Hamiltonian has a Mirror symmetry along the $k_x$-direction, in addition to the    time-reversal symmetry. Therefore, the eigenvalues obey the condition that $E_{k_x}=E_{-k_x}$, which restricts the lowest order term to be quadratic in $k_x$, which is indeed the full calculation suggests. 

For this edge the $k_x$ remains a good quantum number of the eigenstate, while the $k_y \rightarrow i \frac{\partial}{\partial y}$. Given the condition that the wavefunction must die as $y \rightarrow \infty$, we take the trial wavefunction as
\be
\Psi_{k_x}(y)=\begin{pmatrix}
c_{1k_x} \\
c_{2k_x} \\
c_{3k_x} \\
c_{4k_x} \\
\end{pmatrix} e^{- \lambda y},
\hphantom{} \\
\label{edge_wf}
\ee
where $c_{ik_x}$ and $\lambda$ are to be evaluated. We find that the calculation is dramatically simplified if we introduce an anisotropic term to the Dirac mass term as $\Omega_{k}=\frac{\Omega}{2} + B(k_x^2+k_y^2)$, and the final result is obtained with $B\rightarrow 0$. Since this additional term contains quadratic momentum dependence, this does not change the bulk topology. Since $y=0$ axis is set to be the edge, we assume the system for $y>0$ is a non-trivial insulator (having positive Dirac mass, i.e., $\Omega_{k}>0$), while that for $y<0$ is a trivial insulator ($\Omega_{k}<0$). Using the expression for our $4 \times 4$ Hamiltonian, we solve the Schr\''odinger's equation for $y>0$, and find the following coupled equations. (The kinetic term only gives the plane wave solution, it is dropped out in the calculation of the edge states) :
\begin{equation} \label{eq: 1}
\begin{split}
\bigg \{\frac{\Omega}{2} + B (k_x^2 - \lambda^2) - E \bigg\}c_1 -i \alpha_R k_x c_2 - t e^{-\lambda b}c_3\ &=\ 0\\
i \alpha_R k_x c_1 + \bigg\{-\frac{\Omega}{2} - B (k_x^2 - \lambda^2)- E\bigg\}c_2 - t e^{-\lambda b}c_4\ &=\ 0\\
-t e^{\lambda b}c_1 + \bigg\{-\frac{\Omega}{2}  - B (k_x^2 - \lambda^2)- E\bigg\}c_3 + i \alpha_R k_x c_4\ &=\ 0 \\
-t e^{\lambda b}c_2 -i \alpha_R k_x c_3 + \bigg\{\frac{\Omega}{2} + B (k_x^2 - \lambda^2 - E)\bigg\}c_4\ &=\ 0, \\
\end{split}
\end{equation}
where $E$ is the corresponding eigenvalue. For $y<0$ we do the substitution on the mass term (along with the quadratic term) $M(p) \rightarrow -M(p)$ and the trial solutions are :
$\begin{pmatrix}
c_1 \\
c_2 \\
c_3 \\
c_4 \\
\end{pmatrix} e^{ \lambda y}$ \hphantom{} \\
For $y<0$ the derivatives are with respect to $-y$. \hphantom{} \\
We solve the four coupled equations for both cases of $y<0$ and $y>0$ separately and then match the wavefunction at $y=0$, to get the following dispersion and decay length:
\begin{equation}
 \begin{split}
 E_y\ &=\ \pm \bigg(t + \frac{\alpha_R^2}{2 t}\ k_x^2 -\frac{\alpha_R^4 }{8 t^3}\ k_x^4\bigg) \\
 \text{lim}\ & B \rightarrow 0\,, \ \ \lambda \rightarrow \sqrt{\frac{\Omega}{2 B}} 
 \end{split}
\end{equation}

Thus, the edge state decays very fast in the limit $ B \rightarrow 0$. However, the dispersion is not affected. This result matches exactly with the numerical results when we see the dispersion for two 1D
channels, where the gap is of the order $t$. The apparent gap in the edge state by $t$ is an artifact arising due to finite size effect, which disappears as the number of SO wires is increased, gradually reducing the hybridization between the two edges. This result is confirmed by numerical calculation as shown in the main text.  

\subsubsection{Edge parallel to y-axis}
The edge state behaves differently for the edge perpendicular to the SOC wires, i.e., parallel to the $y$-direction. Along this edge, the boundary state is topologically protected but gapped. We have chosen the hybridization between the `A' and `B' wires lying along the $+y$-direction to be finite, and the set the hybridization ($t^{\prime}$) along $-y$ to be zero (the idea is to have these two hybridization to be different so that the imaginary term in the net hybridization survives). This is the origin of a gap in the edge state along this edge. As $t^{\prime}$ is turned on slowly we find that the gap disappears. 

When the edge is parallel to the $y$-axis, the eigenvalue and eigenstates are functions of $k_y$. However, we find that for the energy eigenvalues to be real, the decay length $1/\lambda$ has to be imaginary. This gives {\it only} standing wave solutions along the $x$-direction, and the corresponding wavefunction of have the form: 
\be
\Psi_{k_y}(x)=
\begin{pmatrix}
c_{1k_y} \\
c_{2k_y} \\
c_{3k_y}\\
c_{4k_y} \\
\end{pmatrix} (e^{i \lambda x}-e^{-i \lambda x}). \hphantom{} \\
\ee
Following the aforementioned procedure used for the other edge, we arrive at the energy values: 
\beq
E_{x}&=&\pm \alpha _R \sqrt{\frac{\Omega}{2 B}},\\
\lambda &\rightarrow& \sqrt{\frac{\Omega}{2 B}}. 
\eeq
The apparent divergence of $E_x$ and $\lambda$ as $B\rightarrow0$ is an artifact of the analytical computation for small system size and converges in the numerical calculations when system size is made large. Hence, the gap has leading order in $\alpha_R$, which satisfies the numerical results in the limit $ B\rightarrow 0$.

\subsection{Time reversal invariance}
Under TR symmetry, not only the momentum and spin are reserved, but the magnetic field is also flipped. For this reason, the Zeeman term in our Hamiltonian is also reversed. In what follows, the TR invariance of our Hamiltonian dictates 
\begin{equation} \label{TRS}
\mathcal{U} \mathcal{H}^*(\vec{k}, \Omega) \mathcal{U}^{-1}\ =\ \mathcal{H}(-\vec{k},-\Omega)
\end{equation}
where $\mathcal{U}$ is the unitary operator acting on the spin basis in the TR operator $\tau$, where $\tau = \mathcal{U} \mathcal{K}$,  and $\mathcal{K}$ is the complex conjugation operator.
For our case, the basis states of our Hamiltonian (Eq.~3 in the main text) are $\{A_{\uparrow},A_{\downarrow},B_{\uparrow},B_{\downarrow}\}$, and hence, the unitary operator is:
 \begin{equation} \label{Umatrix}
 \mathcal{U} = \begin{pmatrix}
 0 & -1&0&0\\
 1&0&0&0\\
 0& 0& 0& -1\\
 0& 0& 1& 0
 \end{pmatrix}
\end{equation}
Using Eq.~(\ref{Umatrix}) in Eq.~(\ref{TRS}), we conclude that the $4 \times 4$ Hamiltonian Eq.~3 (in the main text)  is TR symmetric, and hence each of the eigenvalues are doubly degenerate even in the presence of $\Omega$ and $t$. 


\subsection{Calculation of spin density}
We also calculate the spin density for our Hamiltonian in  Eq.~3 (in the main text) to further ascertain the TR invariance of the system. The three spin operators are $S_{x,y,z} = \mathcal{I}_{2\times 2}\otimes \sigma_{x,y,z}$. We indeed find that the expectation values of the spin operators in all three directions $\langle S_{x,y,z}\rangle = 0$. We explicitly discuss the $\langle S_{z}\rangle$ case here. Eigenstates $|1 \rangle$, $|3 \rangle$ and $|2 \rangle$, $|4 \rangle$ have equal and opposite $S_z$ expectation value: 
$ \langle 1| S_z| 1\rangle =  -\langle 4| S_z| 4\rangle$ and $ \langle 2| S_z| 2\rangle =  -\langle 3| S_z| 3\rangle$, giving $ \langle 1| S_z| 1\rangle + \langle 3| S_z| 3\rangle = -4(1 +\frac{\Omega^2}{t^2})$ and, $ \langle 2| S_z| 2\rangle + \langle 4| S_z| 4\rangle = 4(1 + \frac{\Omega^2}{t^2})$ . Thus, the total $S_z$ expectation value of all the eigenstates of the $4 \times 4$ lattice Hamiltonian vanishes, thus showing that TR invariance is preserved for the $4 \times 4$ Hamiltonian of the set up shown in Fig. (1).

\subsection{Band progression}

\begin{figure*}[t]
\rotatebox[origin=c]{0}{\includegraphics[width=1.99\columnwidth]{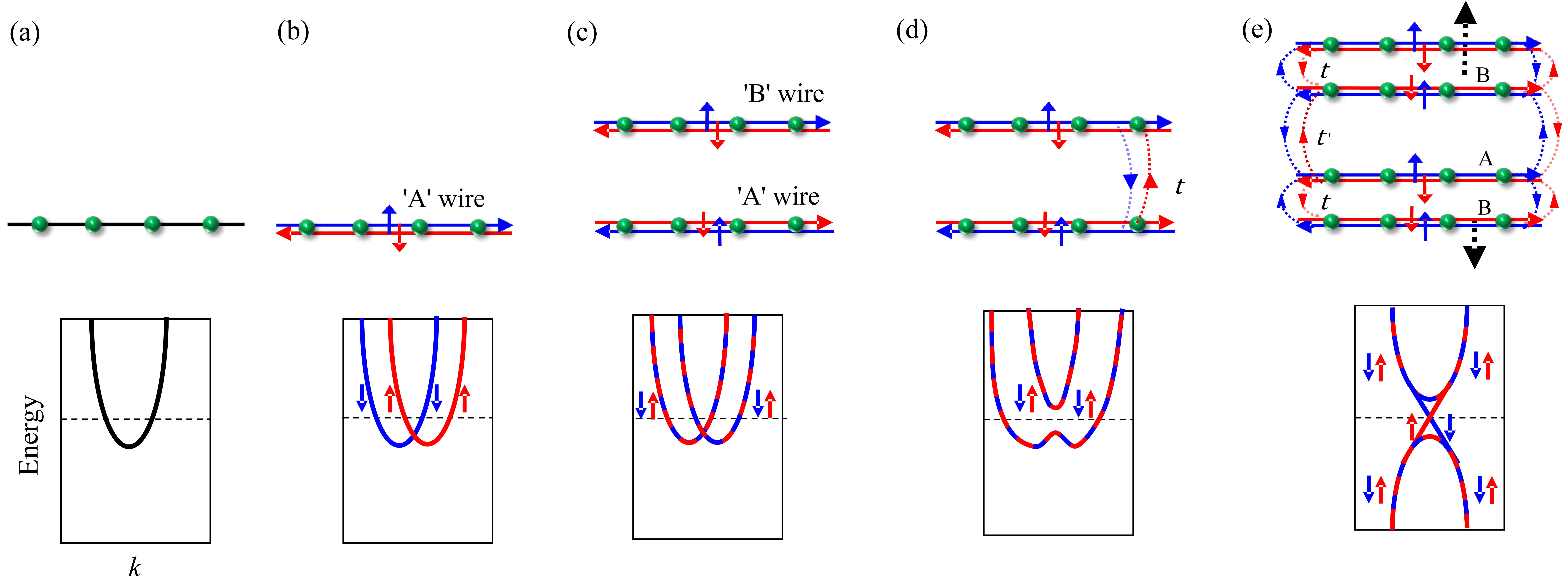}}
\caption{(a) A free-fermionic parabolic band for a single row of atoms without SOC. (b) With 1D SOC but without the Zeeman term ($\Omega=0$), band is spit in the momentum direction, while obeying the Kramer's degeneracy at the $ \Gamma$-point. (c) The split splittings of the `A' and `B' wires are exactly opposite (keeping $\Omega=0$). Therefore, as we overlap the two states, the bands become spin degenerate. (d) As the opposite Zeeman (and/or the quantum tunneling between the wires, $t$) is turned on between the two coupled wires, a band gap opens at the $\Gamma$-point. (e) As an array of coupled SO wires are prepared along the $y$-direction, and as the SOC and quantum tunneling amplitudes are tuned above the respective kinetic energy terms ($\gamma_1$, and $\gamma_2$), the valence band fails to cross the Fermi level, and an full insulating gap opens at all $k$-points. As mentioned in the main text, the system is also associated with $Z_2$ topological invariant due to the property of the SOC reversal. Therefore, along the edge, a spin-polarized edge state arises.}
\label{supp_band}
\end{figure*}

In Fig.~\ref{supp_band} we present how the band structure evolves as we include different terms in the Hamiltonian separately. As ultra-cold atoms are placed in a 1D periodic array, the quantum tunneling between them produces a typical parabolic band. With Rasha-type SOC, the band splits in the momentum space, as illustrated in Fig.~\ref{supp_band}(b). In the `B' wire, we reserve the direction of the SOC, which gives a same band splitting as the previous wire (`A' wire), but the spin expectation value of the bands is reversed. Therefore, the combined setup gives spin-degenerate SOC split bands (as in the case of inversion and TR symmetric systems). As we turn on the Zeeman coupling, but keep it reversed in the two adjacent wires, a band gap opens at the TR invariant $\Gamma$-point even without breaking the TR symmetry. The corresponding band structure is shown in Fig.~\ref{supp_band}(d). With tuning the SOC strength and the inter-wire hopping amplitude the valence band can be pulled back completely below the Fermi level. In this case, as an insulating gap forms, the non-trivial topology ensures a protected metallic surface state.


\begin{thebibliography}{1}
\bibitem{rev1} I. Bloch, Nat. Phys. \textbf{1}, 23 (2005).

\bibitem{rev2} O. Morsch, and  M. Oberthaler, Rev. Mod. Phys.  \textbf{78}, 179  (2006).

\bibitem{rev3} M. Lewenstein, A. Sanpera, V. Ahufinger, B. Damski, A. S. De, U. Sen,  Adv. Phys.  \textbf{56},  243 (2007).

\bibitem{rev4}  A. Derevianko, and  H.  Katori, Rev. Mod. Phys. \textbf{83}, 331 (2011).

\bibitem{Gauge2}  N. Goldman, G. Juzeliunas, P. Ohberg, and  I. B. Spielman, Rep. Prog. Phys. {\bf 77} 126401 (2014).

\bibitem{Gauge_AB}  J. Dalibard, F. Gerbier, G. Juzeliunas, and P. \''Ohberg, Rev. Mod. Phys. {\bf 83}, 1523 (2011).

\bibitem{Gauge} D. Jaksch, and  P. Zoller,  New J. Phys. {\bf 5}, 56 (2003).

\bibitem{HHB} H. Miyake, G. A. Siviloglou, C. J. Kennedy, W. C. Burton, and W. Ketterle, Phys. Rev. Lett. {\bf 111}, 185302 (2013).

\bibitem{lin} Y. J. Lin,  K. J. Garcia, and I. B. Spielman, Nature {\bf 471}, 83 (2011).

\bibitem{jotzu}  G. Jotzu,	M. Messer,	R. Desbuquois, M. Lebrat,	T. Uehlinger, D. Greif, and T. Esslinger,  Nature  {\bf 515}, 237 (2014).

\bibitem{Goldman} N. Goldman, I. Satija, P. Nikolic, A. Bermudez, M. A. Martin-Delgado, M. Lewenstein, and I. B. Spielman, Phys. Rev. Lett. {\bf 105}, 255302 (2010).

\bibitem{aidelsburger} M. Aidelsburger, M. Atala, M. Lohse, J. T. Barreiro, B. Paredes and I. Bloch, Phys. Rev. Lett. {\bf 111}, 185301 (2013).

\bibitem{mancini} M. Mancini, G. Pagano, G. Cappellini, L. Livi, M. Rider, J. Catani, C. Sias, P. Zoller, M. Inguscio, M. Dalmonte, and L. Fallani, Science. {\bf 349} 1510 (2015). 

\bibitem{Stuhl}B. K. Stuhl, H.I. Lu, L. M. Aycock, D. Genkina, and I. B. Spielman, Science {\bf 349}, 1514 (2015).

\bibitem{TIreviewTD}  A. Bansil,  H. Lin, and  T.  Das, Rev. Mod. Phys. Rev. Mod. Phys. {\bf 88}, 021004 (2016) .

\bibitem{TIreviewCK} M. Z. Hasan, and  C.L. Kane,  Rev. Mod. Phys.~{\bf 82}, 3045 (2010).

\bibitem{TIreviewSCZ} X. L. Qi, and  S.C.  Zhang, Rev. Mod. Phys.~{\bf 83}, 1057 (2011).

\bibitem{bliokh}  K. Y. Bliokh,  D. Smirnova, and F. Nori, Science {\bf 348}  1448 (2015). 

\bibitem{SOC2D} L. Huang, Z. Meng, P. Wang, P. Peng, S.L. Zhang, L. Chen, D. Li, Q. Zhou, and J. Zhang,   arXiv:1506.02861.

\bibitem{SOC2D2} Z. Wu, L. Zhang, W. Sun, X.T. Xu, B.Z. Wang, S.C. Ji, Y. Deng, S. Chen, X.J. Liu, and J.W. Pan,  arXiv:1511.08170 .

\bibitem{bhz}  B. A. Bernevig,  T. L. Hughes,  and  S.C. Zhang, Science {\bf 314} , 1757  (2006). 

\bibitem{kane} C. L. Kane, and  E. J. Mele,  Phys. Rev. Lett. {\bf 95}, 146802 (2005).


\bibitem{footnote_setup} Two SO wires are not required to be placed on the same plane. If they are placed on different planes (along the $z$-direction) to facilitate preparations, the inter-wire hopping in Hamiltonian (Eq.~\ref{Hamp}) would still remain the same as long as the distance ($b$) along the $y$ direction is kept to be the same.  

\bibitem{footnote_complexhopping} In the real-space, the hoppings from `B' to the top and the bottom `A' wires are taken to be different ($t\ne t^{\prime}$) without any complex phase associated with it. This staggered hopping naturally leads to a complex Bloch in the momentum space, according to the SSH model,\cite{ssh1} and serves our purpose. This can be achieved by tuning the distance between the two adjacent wires to be slightly different, as illustrated in Fig.~\ref{fig1}(b).


\bibitem{ssh1} W. P. Su,  J. R. Schrieffer, and  A. J. Heeger, Phys. Rev. Lett. {\bf 42}, 1698 (1979).

\bibitem{fu} L. Fu,  and  C.L.  Kane,  Phys. Rev. B {\bf 74} , 195312 (2006).

\bibitem{soluyanov}  A. A. Soluyanov, and  D. Vanderbilt, Phys. Rev. B {\bf 83} , 035108 (2011). 

\bibitem{yu} R. Yu, X. L. Qi, A. Bernevig, Z. Fang and X. Dai, Phys. Rev. B. {\bf 84} 075119 (2011).

\bibitem{Roth}  A. Roth, C. Brüne, H. Buhmann, L. W. Molenkamp, J. Maciejko, X.L. Qi, and S.C. Zhang, Science {\bf 325}, 294 (2009).

\bibitem{QW1} I. Barke , F. Zheng, T. K. Rugheimer,  and  F. J. Himpsel,  Phys. Rev. Lett. {\bf 97}, 226405 (2006).

\bibitem{QW2}  C. Brand, 	H. Pfnür, G. Landolt, S. Muff, J. H. Dil, T. Das, and C. Tegenkamp, Nat. Commun. {\bf 6}, 8118  (2015).

\bibitem{NCTD} T. Das, and   A. V.  Balatsky, Nat. Commun. {\bf 4}, 1972 (2013). 

\bibitem{fujii}K. Fujii, K. Higashida, R. Kato, and Y. Wada, arXiv:quant-ph/0307066v2 .

\bibitem{sanchez}B. N. Sanchez, and  T. Brandes, {Ann. Phys.} {\bf 13}, 569-594 (2004).

\bibitem{brion}E. Brion, L. H. Pedersen, and K. Molmer,  arXiv:quant-ph/0610056.

\bibitem{shen}S. Q. Shen, Topological Insulators: Dirac Equation in Condensed Matters. (Springer 2012).


\end{thebibliography}

\begin{thebibliography}{1}

\bibitem{lin} Y. J. Lin, K. Jimenez-Garcia, and I. B.  Spielman, {Nature Lett}. {\bf 471}, 83 (2011).

\bibitem{fujii}K. Fujii, K. Higashida, R. Kato, and Y. Wada, arXiv:quant-ph/0307066v2 .

\bibitem{sanchez}B. N. Sanchez, and  T. Brandes, {Ann. Phys.} {\bf 13}, 569-594 (2004).

\bibitem{TIreviewCK}M. Z. Hasan, and C. L.  Kane, {Rev. Mod. Phys}.~{\bf 82}, 3045 (2010).

\bibitem{TIreviewSCZ}X. L. Qi, and S.C.  Zhang, {Rev. Mod. Phys}.~{\bf 83}, 1057 (2011).

\bibitem{TIreviewTD}A. Bansil, H. Lin, and  T. Das, {Rev. Mod. Phys.} (to be published).

\bibitem{bhz}B. A. Bernevig, T. L Hughes, and S. C. Zhang, {Science} {\bf 314} , 1757  (2006). 

\bibitem{kane}C. L. Kane, and E. J. Mele, {Phys. Rev. Lett}. {\bf 95}, 146802 (2005).

\bibitem{brion}E. Brion, L. H. Pedersen, and K. Molmer,  arXiv:quant-ph/0610056.

\bibitem{shen}S. Q. Shen, Topological Insulators: Dirac Equation in Condensed Matters. (Springer 2012).


\end{thebibliography}
\end{document}